\documentclass[slac_one]{revtex4}
\usepackage{graphicx}
\usepackage{fancyhdr}
\def\nyu{Physics Department, New York University\\
4 Washington Place, New York, NY 10003, USA}

\def\MET{\textsl{E}$_{\mbox{\footnotesize{T}}}^{\mbox{\footnotesize{miss}}}$}

\def\ET{\textsl{E}$_{\mbox{\footnotesize{T}}}$}

\begin{document}

\title{Performance of the ATLAS Tau and Missing Energy triggers with 7 TeV proton proton collisions at the LHC} 

%

\author{L. Hooft van Huysduynen \footnote{For the ATLAS Collaboration}}
\affiliation{\nyu}

\begin{abstract}
\noindent \textbf{Abstract.} A study of the performance of the ATLAS tau and missing energy triggers with data collected in spring 2010 at $\sqrt{s} = 7$ TeV proton-proton collisions at the Large Hadron Collider (LHC) 
is presented.

A comparison was performed between data and Monte Carlo simulations for the tau and missing transverse energy triggers. As well as a comparison between missing transverse energy trigger quantities and their offline reconstructed counterparts.

Tau trigger results compare well with predictions from Monte Carlo simulations. Slight deviations are observed for tau shower shape quantities. Possible sources contributing to the discrepancy such as the
simulation of the underlying event are currently being studied.

The missing transverse energy reconstructed by the Event Filter is well correlated with the offline result.  In addition, there is good agreement between the results obtained with collision data and Monte Carlo simulations.
\end{abstract}

\maketitle

\thispagestyle{fancy}

\section{Introduction}
\noindent The ATLAS tau and missing transverse energy performance was studied using the 7 TeV collisions recorded at the LHC in the spring of 2010.
Both are important signatures for the study of the standard model and in the search for new physics.
To determine the performance of the triggers a comparison was made between the MC simulation and the data collected. The results obtained are presented in this note.

\section{ATLAS Missing Energy and Tau Trigger}
\noindent The ATLAS detector deploys a three-level trigger scheme to select potentially interesting physics events and to reduce the amount of data events to be reconstructed and stored. \cite{ATLAS}. 
Level 1 (L1) is implemented in hardware using fast custom built electronics. Level 2 (L2) and Event-Filter (EF) are software based, running their algorithms on a computer farm.

The tau L1 hardware trigger is based on electromagnetic (EM) and hadronic (HAD) calorimeter trigger towers (TTs) of size $0.1 \times 0.1$ 
in $(\eta,\phi)$. Four EM and HAD TTs are used to select a local \ET$\ $ maximum Region-of-Interest (RoI). Narrowness, isolation and 
low track multiplicity of the tau jet are used to distinguish from the multijet background at L2 and EF.

The missing transverse energy trigger requires the magnitude of the vector sum of all transverse energies to exceed some threshold. 
At L1 only calorimeter information from TTs is used. 
At L2, the L1 result is reused with an additional correction for muons reconstructed at L2, this correction is currently is not included in the trigger decision. At EF, the missing transverse energy (\MET) is recomputed using the full granularity of the detector.

\section{Tau Trigger Performance}
\noindent A pure sample of real hadronic tau decays is not available yet. 
Therefore tau-like QCD events are used to study the performance of the tau trigger in data \cite{TAUCONFNOTE}. All Figures shown here compare minimum bias collision data with the expected results from MC simulations.

\begin{figure}[h!]
  \begin{minipage}[c]{.49\linewidth}
    \begin{center}
      \includegraphics[width=3.1in]{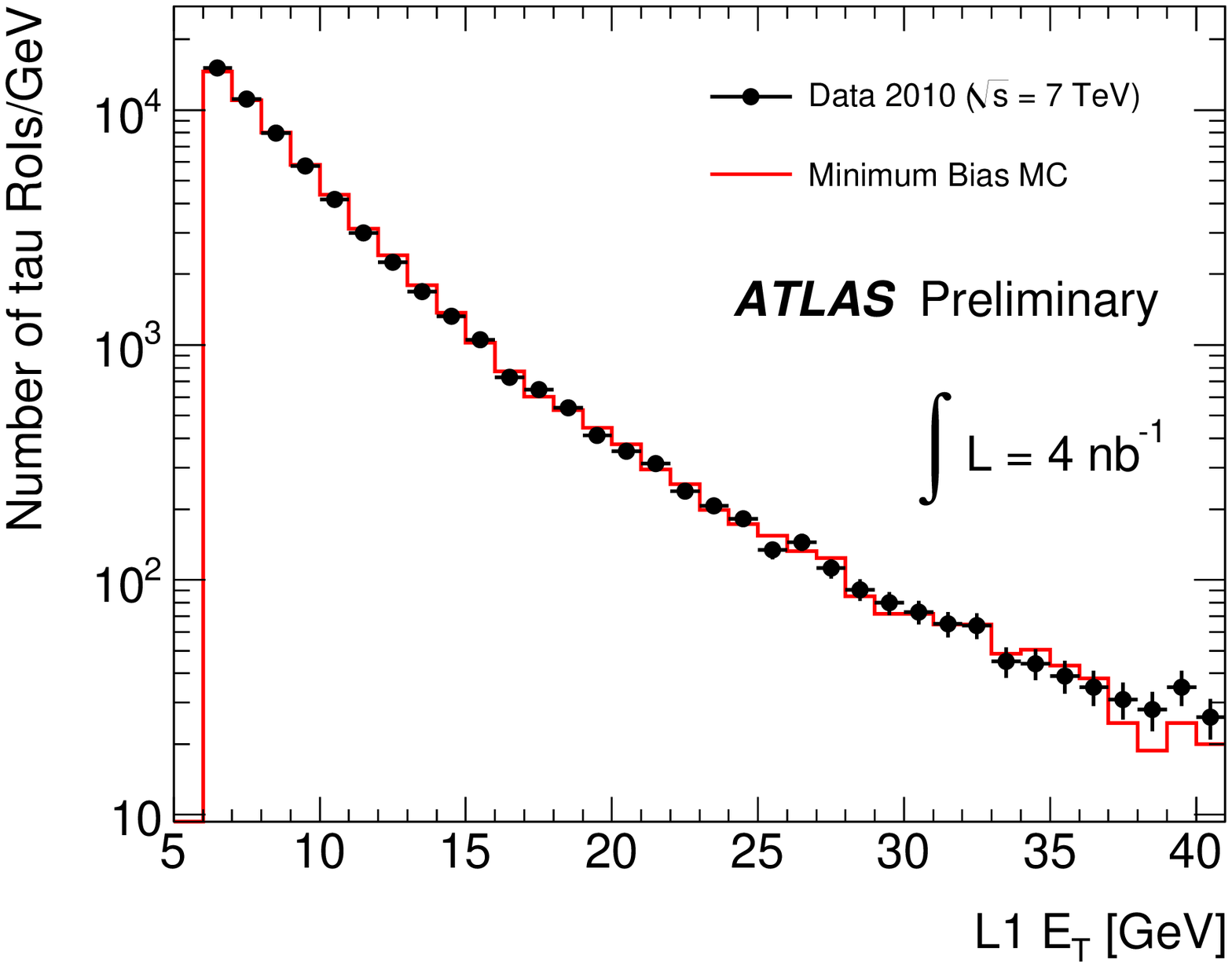}
    \end{center}
  \end{minipage}\hfill
  \begin{minipage}[c]{.51\linewidth}
    \begin{center}
      \caption{Comparison of the L1 tau candidate $E_{T}$ distribution for 7 TeV data and minimum bias MC simulation. The cut-off at 6 GeV corresponds to the lowest L1 threshold.}
      \label{fig:TAU_ET}
    \end{center}
  \end{minipage}
  \begin{minipage}[c]{.49\linewidth}
    \begin{center}
      \includegraphics[width=3.1in]{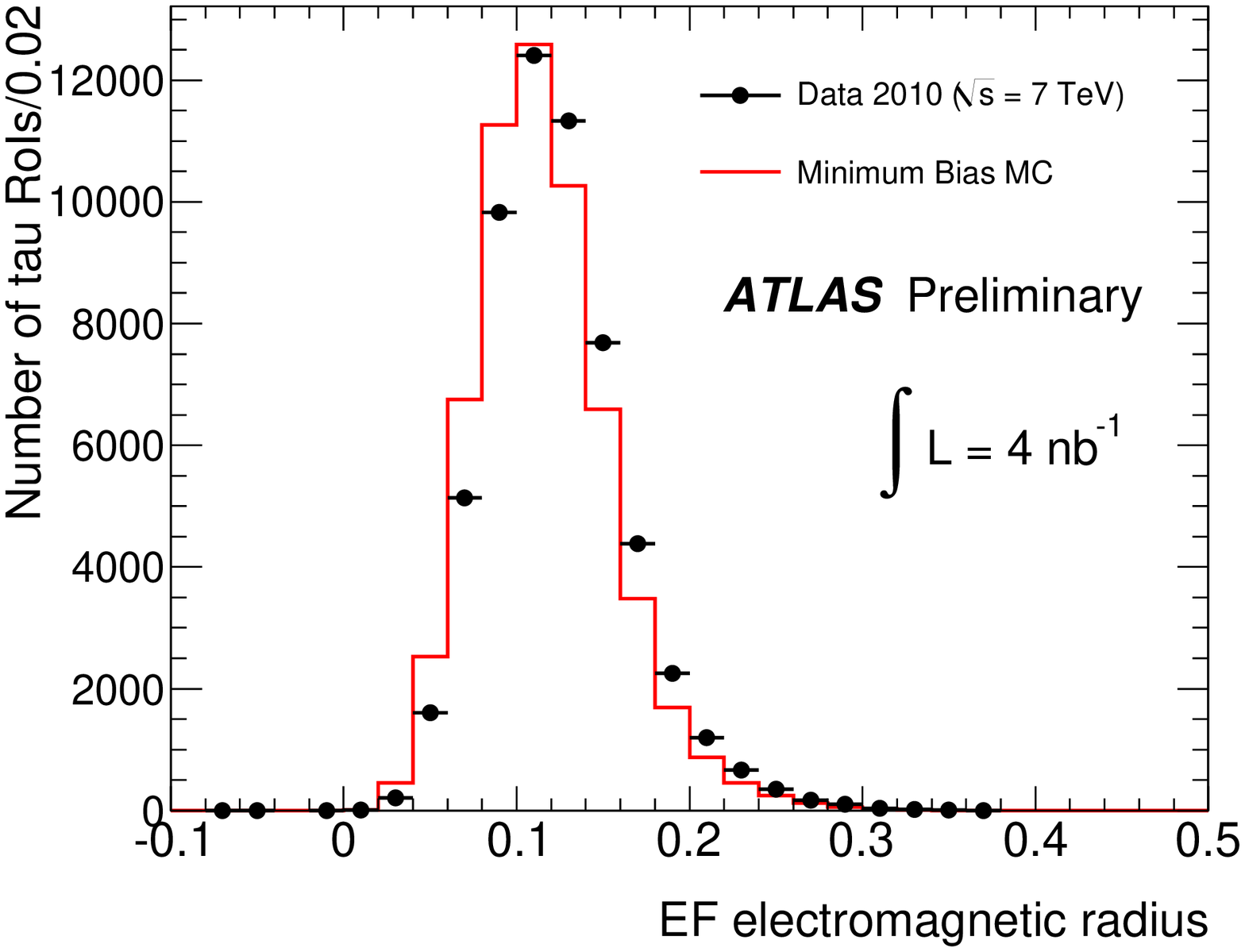}
      \caption{Comparison of the tau candidate EM radius distribution at EF for 7 TeV data and MC. MC has been normalised to the number of entries of the data histogram.}
        \label{fig:TAU_EMRADIUS}
    \end{center}
  \end{minipage}\hfill
  \begin{minipage}[c]{.49\linewidth}
    \begin{center}
      \includegraphics[width=3.1in]{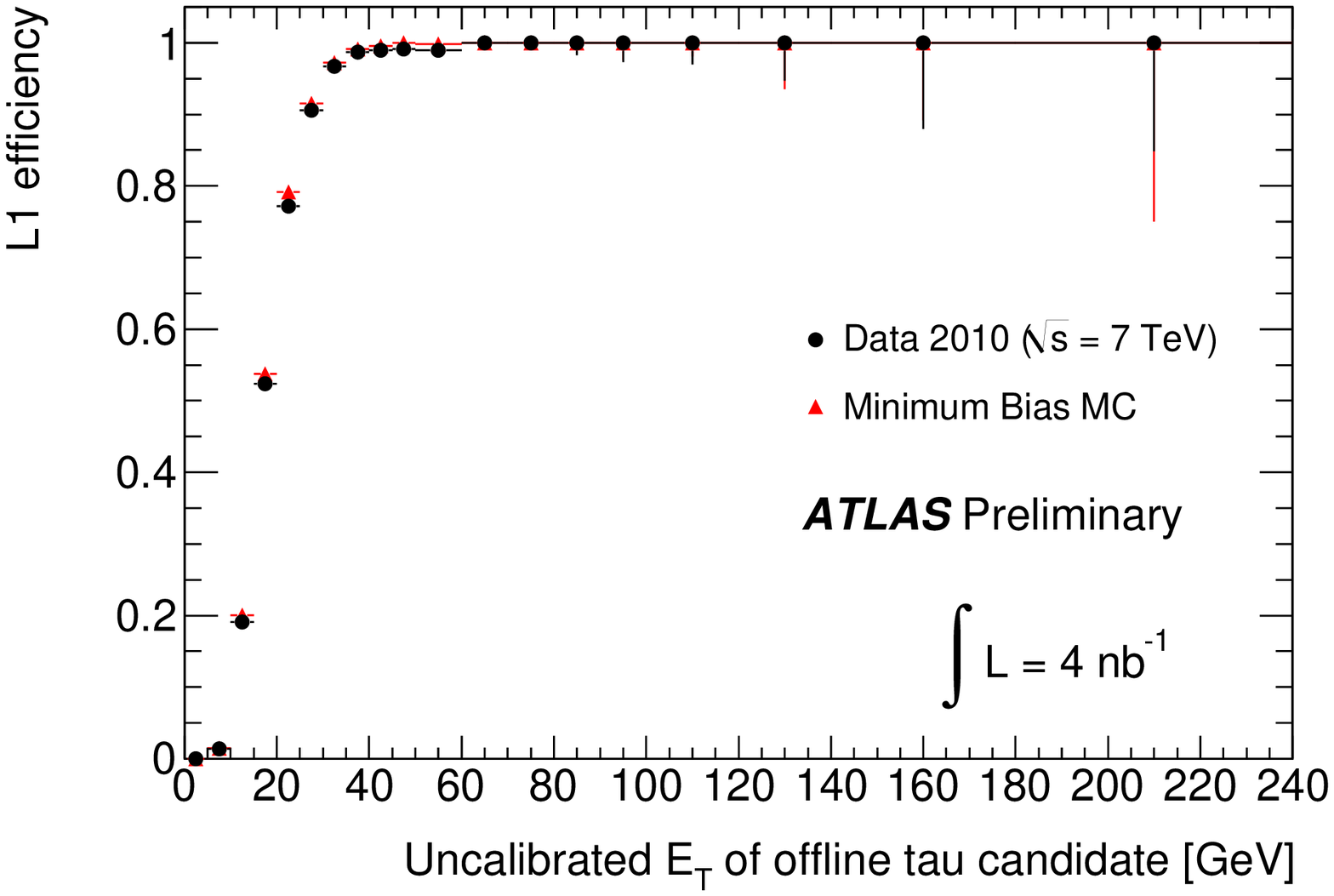}
      \caption{Fraction of the offline tau candidates matched to a L1 trigger object with $E_{T}$ larger than 5 GeV as a function of the $E_{T}$ of the offline tau candidate.}\label{fig:TAU_TURNON}
    \end{center}
  \end{minipage}
\end{figure}

Figure \ref{fig:TAU_ET} shows the \ET$\ $ distribution of tau candidates at L1, displaying a good agreement between collision data results and MC simulation.

Figure \ref{fig:TAU_EMRADIUS} shows the energy-weighted EM radius at EF, which is a measure of the shower shape. Possible sources that can contribute to the discrepancy such as the simulation of the underlying event are currently being studied.

The turn-on curve measures the probability for offline reconstructed taus to be selected at L1 as a function of the \ET$\ $ of the offline tau candidate. Figure \ref{fig:TAU_TURNON} shows that predictions from MC simulations match the collected data with high precision.

\section{Missing Energy Trigger Performance}
\begin{figure}[h!]
  \begin{minipage}[c]{.49\linewidth}
    \begin{center}      
      \includegraphics[width=3.in]{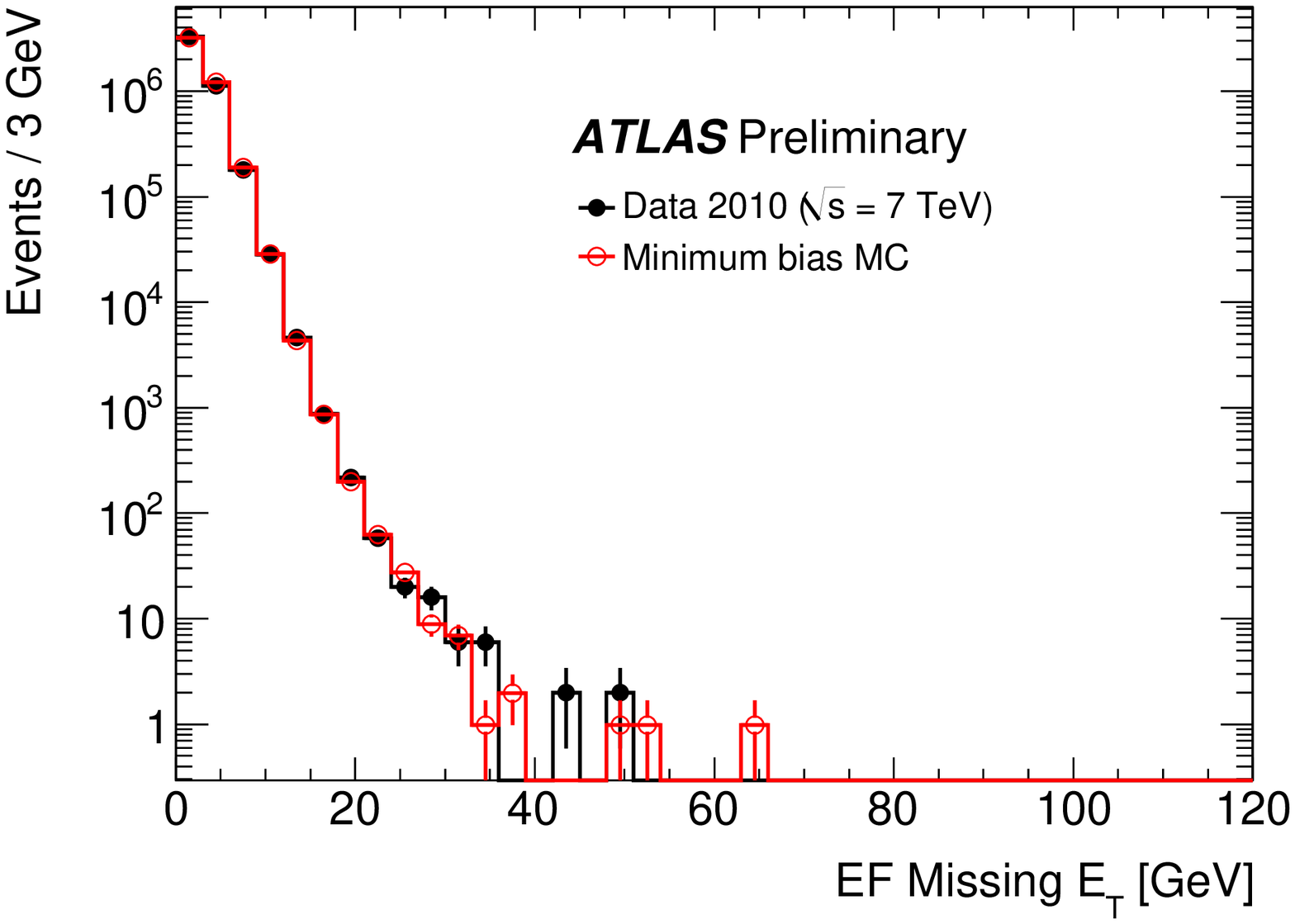}
    \end{center}
  \end{minipage}\hfill
  \begin{minipage}[c]{.51\linewidth}
    \begin{center}
      \caption{EF \MET$\ $ distribution for real 2010 7 TeV data and MC events, after applying jet clean-up cuts. The MC distribution has been normalized to the data distribution.}
      \label{fig:MET_EF}
    \end{center}
  \end{minipage}
  \begin{minipage}[c]{0.49\linewidth}
    \begin{center}
      \includegraphics[width=3.5in]{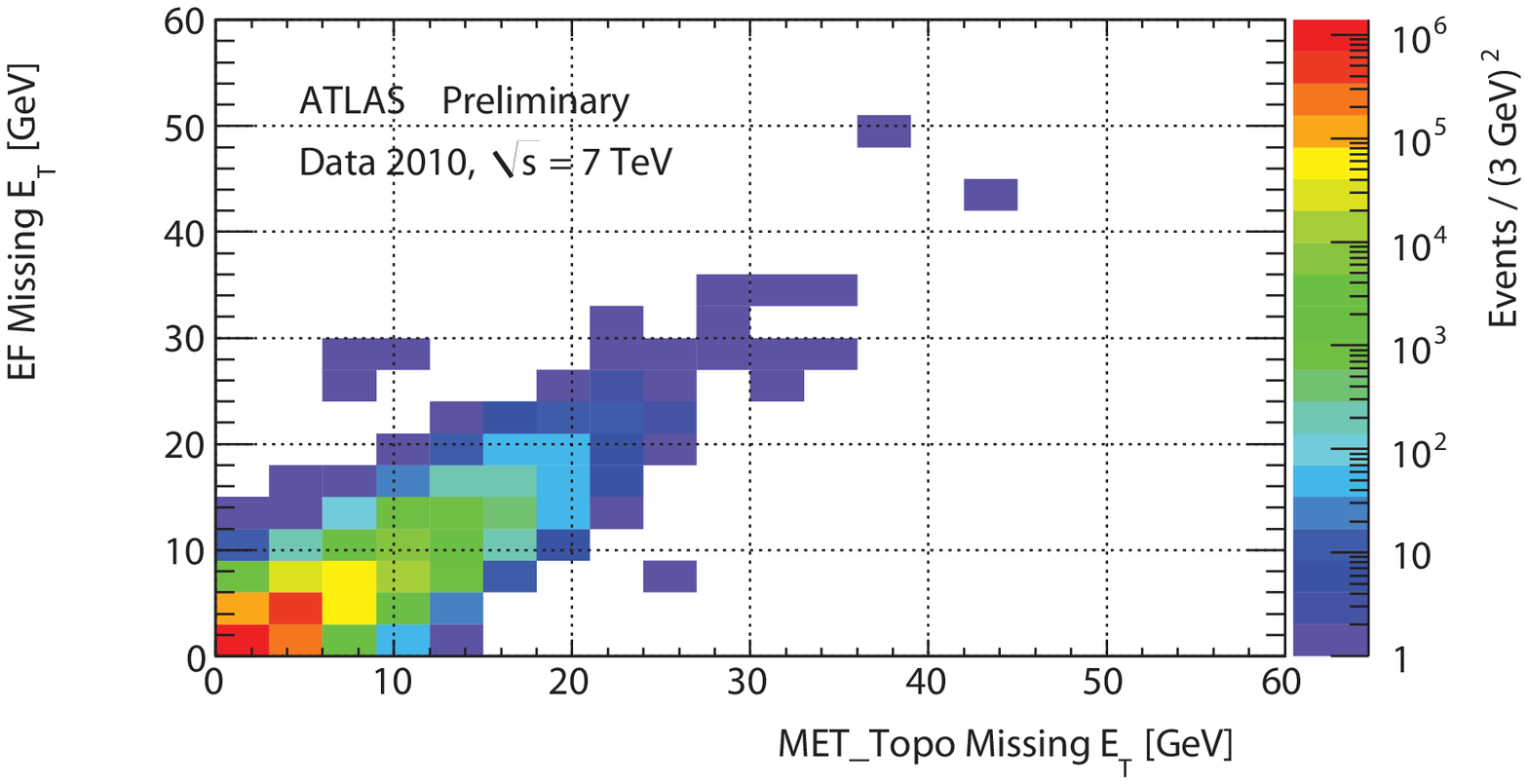}
      \caption{Correlation between the \MET$\ $ measured at EF and offline, after the jet clean-up cuts.}
      \label{fig:MET_TOPO_L1vsEF}
    \end{center}  
  \end{minipage}\hfill
  \begin{minipage}[c]{0.49\linewidth}
    \begin{center}
      \includegraphics[width=3.2in]{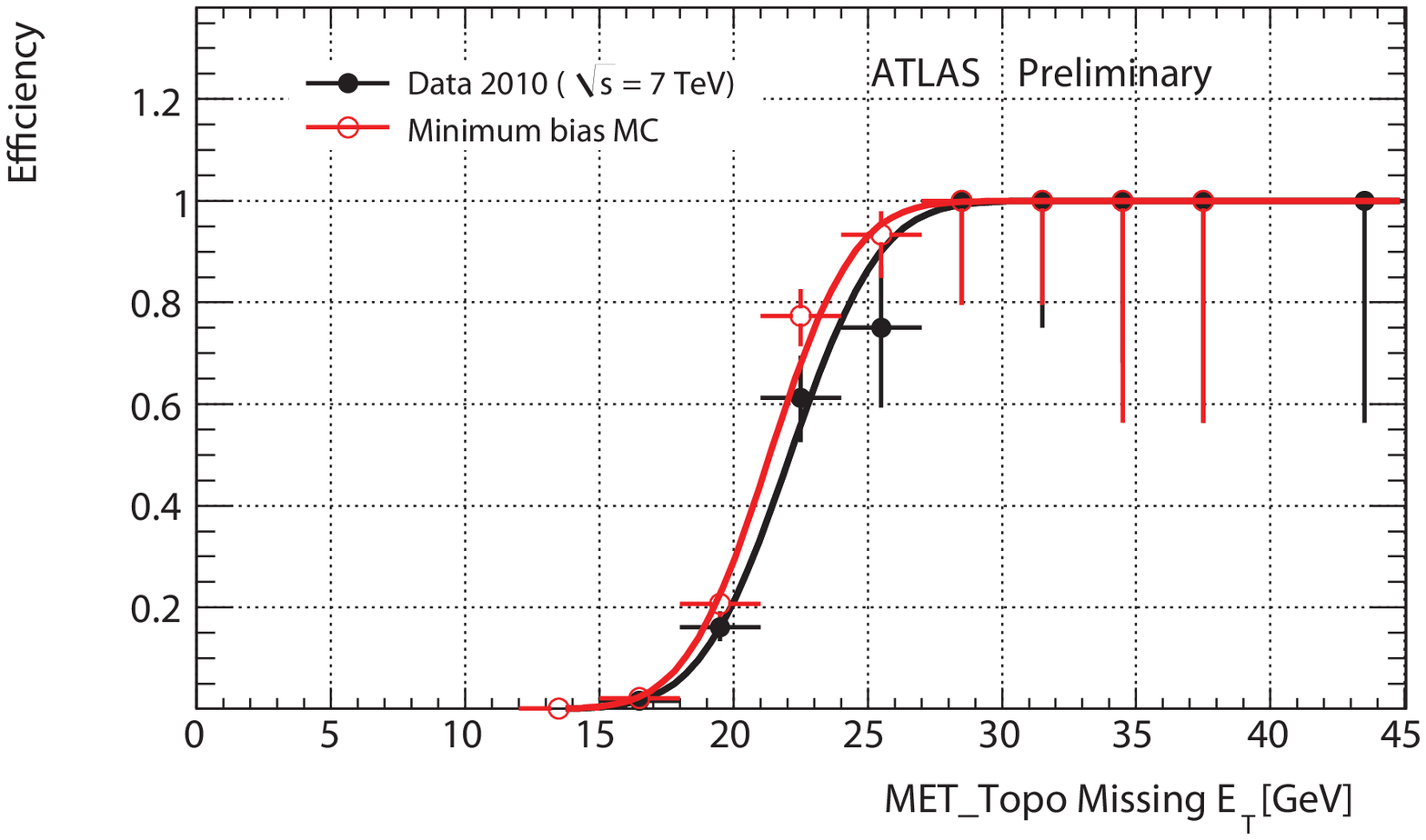}
      \caption{Turn-on curves for \MET$\ $ thresholds of 20 GeV at EF for collision candidates and simulated events.}
      \label{fig:MET_TURNON}
    \end{center}  
  \end{minipage}
\end{figure}

\noindent Good agreement is shown in Figure \ref{fig:MET_EF} between the EF \MET$\ $ distributions for real collision candidate data, after applying jet clean-up cuts, and MC events from minimum bias events  \cite{METCONFNOTE}.

Figure \ref{fig:MET_TOPO_L1vsEF} shows the correlation between measured \MET$\ $ at EF and its offline reconstructed counterpart for collision candidates after applying jet clean-up cuts. 
A clear linear relationship exists between the trigger and offline reconstruction.

The MC EF turn-on curve (Figure \ref{fig:MET_TURNON}) closely describes real collision candidate data results, after applying jet clean-up cuts, and the 100$\%$ efficiency plateau is reached close to the trigger threshold\\

\section{Conclusions}
\noindent Performance studies demonstrate that the tau trigger can be safely used for physics data taking, though minor descrepancies between the shape variables 
measured with real and MC events remain to be fully understood.

The calorimeter-based missing energy trigger has been validated and it is currently enabled, rejecting events at L1, L2 and EF, and is used in various physics analyses. Work is in progress to 
commission a muon correction term at L2 and EF.

\end{document}